\documentclass[aps,preprint,showpacs]{revtex4-1}

\usepackage{amssymb}
\usepackage{mathrsfs}
\usepackage{lipsum}
\usepackage{graphicx}
\usepackage[caption=false]{subfig}
\usepackage{mathtools}
\usepackage{epstopdf}
\usepackage{xcolor}
\usepackage{color}
\definecolor{atrovirens}{RGB}{100,160,0}
\usepackage[colorlinks, linkcolor=red, anchorcolor=green, citecolor=blue, urlcolor=atrovirens]{hyperref}
\DeclareGraphicsExtensions{.pdf,.jpg,.png,.eps}
\usepackage[toc,page,title,titletoc,header]{appendix}

\begin{document}
	
\title{A nearly optimal and robust protocol for nonlinear phase estimation using coherent states}

\author{Jian-Dong Zhang}
\affiliation{School of Physics, Harbin Institute of Technology, Harbin 150001, China}
\author{Zi-Jing Zhang}
\email[]{zhangzijing@hit.edu.cn}
\affiliation{School of Physics, Harbin Institute of Technology, Harbin 150001, China}
\author{Jun-Yan Hu} 
\affiliation{School of Physics, Harbin Institute of Technology, Harbin 150001, China}
\author{Long-Zhu~Cen} 
\affiliation{School of Physics, Harbin Institute of Technology, Harbin 150001, China}
\author{Yi-Fei Sun} 
\affiliation{School of Physics, Harbin Institute of Technology, Harbin 150001, China}
\author{Chen-Fei Jin}
\email[]{jinchenfei@hit.edu.cn}
\affiliation{School of Physics, Harbin Institute of Technology, Harbin 150001, China}
\author{Yuan Zhao}
\email[]{zhaoyuan@hit.edu.cn}
\affiliation{School of Physics, Harbin Institute of Technology, Harbin 150001, China}

\date{\today}
	
\begin{abstract}
	We propose a protocol for the second-order nonlinear phase estimation with a coherent state as input and balanced homodyne detection as measurement strategy.
	The sensitivity is sub-Heisenberg limit, which scales as $N^{-3/2}$ for $N$ photons on average.
	By ruling out hidden resources in quantum Fisher information, the fundamental sensitivity limit is recalculated and compared to the optimal sensitivity of our protocol.
	In addition, we investigate the effect of photon loss on sensitivity, and discuss the robustness of measurement strategy.
	The results indicate that our protocol is nearly optimal and robust. 
\end{abstract}

\maketitle

\section{Introduction}

	\label{s1}
	Within the past two decades, quantum technologies have developed at an unheard-of rate. There are a great deal of revolutionary progresses in proof-of-principle experiments, offering an insight into world from microscopic view.
	As a momentous component of quantum technologies, quantum metrology \cite{PhysRevLett.96.010401,TAYLOR20161} is a science that exploits exotic quantum resources to enhance estimation precision of physical quantities.
	In this regard, quantum-enhanced interferometers come across as a suitable tool and play a paramount role.
	They work by mapping a small variation of interest onto an unknown relative phase shift between the two arms and by estimating this phase.

	In recent years, linear phase estimation has received a boost with an influx of demands from the rapidly developing field of quantum information processing.
	Exotic input states and novel measurement strategies have aroused wide interests, so long as they are capable of breaking the shot-noise limit or Rayleigh diffraction limit.
	Among these inputs, two-mode squeezed vacuum and entangled coherent states are probably the greatest candidates, which can even surpass the Heisenberg limit.
	Regarding measurement strategies, parity, on-off, and projective measurements have shown extraordinary performances—optimal or robust or both—in different scenarios.

	As another important element, nonlinear processes also are of vital significance.
	Many of the exotic quantum resources are produced during nonlinear light-matter interactions, e.g., preparations for squeezed and superposition states \cite{PhysRevLett.106.013603,PhysRevLett.101.233605}.
	Related to this, nonlinear phase estimation has also gained lots of attention \cite{PhysRevLett.105.010403,PhysRevA.90.063838,PhysRevA.86.043828,PhysRevA.66.013804,PhysRevA.34.3974,PhysRevA.88.013817,PhysRevA.92.022104}.
	However, most of these protocols only provide the sensitivity limits calculated from the quantum Fisher information (QFI).
	That is, the optimal measurement strategy saturating the QFI is not provided.
	Furthermore, the QFI-only calculation may be a loosen lower bound, since it is on the cards that some hidden resources dilute the tightness.
	Hence, there are some gaps to be filled in this respect, and one needs to study those protocols containing specific measurement strategy.
	To this end, here we propose a estimation protocol for nonlinear phase shifts through the use of coherent states and balanced homodyne detection.
	The QFI is recalculated via the phase-averaging approach, in which any hidden resources are eliminated \cite{PhysRevA.85.011801,PhysRevA.96.052118,PhysRevA.99.042122}.

	\section{Estimation protocol}
	\label{s2}
\begin{figure}[htbp]
		\centering\includegraphics[width=0.7\textwidth]{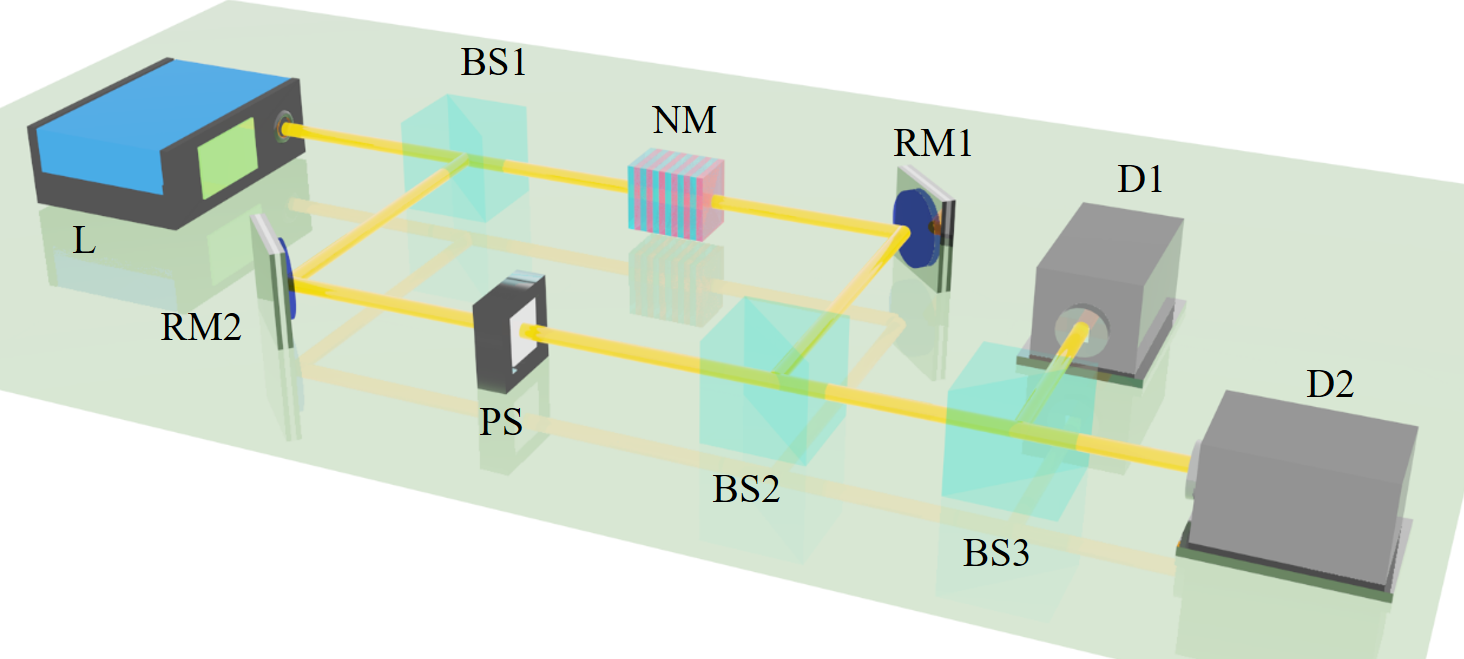}
		\caption{Schematic of estimation protocol for nonlinear phase shifts. The abbreviations are defined as follows: L, laser; BS, beam splitter; RM, reflection mirror; NM, nonlinear medium; PS, phase shifter; D, detector.}
		\label{f1}
	\end{figure}

	We start off with the introduction of our estimation protocol.
	Consider a Mach-Zehnder interferometer as depicted in Fig. \ref{f1}, a nonlinear medium and a phase shifter are inserted into its two paths.
	The clockwise and counterclockwise paths are labeled as spatial modes $A$ and $B$, respectively.
	Throughout this paper, ${\hat a}^\dag$ (${\hat b}^\dag$) and ${\hat a}$ (${\hat b}$) stand for the creation and annihilation operators in mode $A$ ($B$).
	The phase shifter is used to offset the linear phase induced by the nonlinear medium.
	The first input port is fed by a coherent state generated by a laser, and the second one is empty.
	Thus, the input state can be delineated as ${\left| \alpha  \right\rangle _A}{\left| 0 \right\rangle _B}$.	
	Upon leaving the first 50-50 beam splitter, this state goes to
	${\left| {{\alpha }/{{\sqrt 2 }}} \right\rangle _A}{\left| {{{i\alpha }}/{{\sqrt 2 }}} \right\rangle _B}$.

	Without loss of generality, the $k$th-order nonlinear phase operator can be described as ${\hat U_k}\left( \varphi  \right) =  \exp [ {i\varphi {{( {{{\hat a}^\dag }\hat a})}^k}}]$ with respect to nonlinear phase $\varphi$.
	The linear phase $\theta$ is phase difference between the two modes after compensation by the phase shifter.
	For simplicity, we consider the scenario that the linear phase is only in mode $B$; accordingly, the state after experiencing phase shift $\varphi$ becomes
	\begin{equation}
	\left| \psi  \right\rangle  = {\exp [ {i\varphi{{( {{{\hat a}^\dag }\hat a} )}^2}}] \exp ( {i\theta {{\hat b}^\dag }\hat b} )} {\left| {\frac{\alpha }{{\sqrt 2 }}} \right\rangle _A} {\left| {\frac{{i\alpha }}{{\sqrt 2 }}} \right\rangle _B},
	\label{e4}
	\end{equation}

	Finally, this state is incident on the second 50-50 beam splitter, and balanced homodyne detection is performed at the output.
	
	\section{Measurement and estimation in a lossless scenario}
	\label{s3}
	Balanced homodyne detection was originally developed by Yuen and Chan \cite{Yuen:83}.
	It is a process, by which the output state is mixed with a phase-tunable local oscillator, which itself is a coherent state of the same frequency as the input. 
	In Fig. \ref{f1}, the local oscillator is injected into the third 50-50 beam splitter, and is not shown for simplicity.
	This measurement strategy is a standard technique for quantum noise detection by detecting quadrature-phase or quadrature-amplitude. 
	For Gaussian inputs, even without a photon-number-resolving detector, one can utilize this strategy to measure the parity of the output \cite{1367-2630-12-11-113025}.

	Consider the $X$ quadrature of path $B$, the measurement operator can be expressed as ${\hat X_B} = \hat b + {\hat b^\dag }$, and the expectation value of this operator is equal to
	\begin{align}
	\left\langle {{{\hat X}_B}} \right\rangle = \frac{1}{{\sqrt 2 }}\left\langle \psi  \right|[ {( {\hat b + {{\hat b}^\dag }} ) + i( {\hat a - {{\hat a}^\dag }})} ]\left| \psi  \right\rangle.
	\label{e6} 
	\end{align}
	Where the sequitur $\hat U_{\rm BS}^\dag \hat b{{\hat U}_{\rm BS}^{}} = ( {\hat b + i\hat a}) / {\sqrt 2 }$ derived from the Baker-Hausdorff lemma is used, and the operator of 50-50 beam splitter is given by 
	${\hat U_{\rm {BS}}} = \exp [ {i{\pi }( {{{\hat a}^\dag }\hat b + {{\hat b}^\dag }\hat a} )}/4]$.
	
	The expectation value of the first term in Eq. (\ref{e6}) is found to be
	\begin{align}
	\frac{1}{{\sqrt 2 }}\left\langle {\hat b + {{\hat b}^\dag }} \right\rangle  =  - \left| \alpha  \right|\sin \theta, 
	\label{e9}
	\end{align}
	here we assume that the parameter $\alpha$ is a positive real number.
	Regarding the second term, it can be calculated through the use of the lemma $\hat ag( {{{\hat a}^\dag }\hat a} ) = g( {{{\hat a}^\dag }\hat a + 1})\hat a$ \cite{Louisell1973Quantum}.
	We decompose the nonlinear phase term into
	\begin{equation}
	\exp (i\varphi {\hat n_A^2 } ) = \exp [ {i\varphi ({{\hat n}_A^2} - {{{\hat n}_A}}) }]\exp \left( {{{i\varphi\hat n}_A} } \right),
	\label{e11}
	\end{equation}
	where the term $ \exp \left( {{{i\hat n}_A}\varphi } \right)$ is ignored since it denotes a linear phase shift.
	By means of the above lemma, we have
	\begin{align}
	{\hat U}_{\rm PS}^\dag \hat a {\hat U}_{\rm PS}^{} &= \exp \left( {i2\varphi {{\hat n}_A}} \right)\hat a 
	\label{e12}
	\end{align}
	with operator ${\hat U}_{\rm PS}^{} = \exp \left[ {i\varphi\left({{\hat n}_A^2} - {{{\hat n}_A}} \right) } \right]$.
	Further, using the lemma $\left\langle \alpha  \right|\exp \left( {c{{\hat a}^\dag }\hat a} \right)\left| \alpha  \right\rangle   =  \exp [ {\left( {{e^c} - 1} \right)}\left| \alpha \right|^2 ]$ \cite{Louisell1973Quantum},
	we can obtain the expectation value of the second term in Eq. (\ref{e6}),  
	\begin{align} 
	\frac{i}{{\sqrt 2 }}\left\langle {\hat a - {{\hat a}^\dag }} \right\rangle  =  - \left| \alpha  \right|\exp ( { - N{{\sin }^2}\varphi })\sin \left[ {{{\frac{N}{2}}}\sin \left( {2\varphi } \right)} \right]
	\label{e15}
	\end{align}
	with $N={{\left| \alpha  \right|}^2}$ being the mean photon number inside the interferometer.
	Substituting Eq. (\ref{e6}) by Eqs. (\ref{e9}) and (\ref{e15}), the expectation value of $X$ quadrature is obtained
	\begin{equation}
	\left\langle {{{\hat X}_B}} \right\rangle  =  - \left| \alpha  \right|\left\{ {\sin \theta  + \exp ( { - N{{\sin }^2}\varphi })\sin \left[ {{{\frac{N}{2}}}\sin \left( {2\varphi } \right)} \right]} \right\},
	\label{e16}
	\end{equation}
	and that of the square of $X$ quadrature turns out to be
	\begin{align}
	\nonumber \left\langle {\hat X_B^2} \right\rangle  =& N- \frac{N}{2} \left\{\cos \left( {2\theta } \right) - \exp ({ - 2N{{\sin }^2}\varphi })\cos \left[ {N\sin \left( {2\varphi } \right)} \right] \right\}+ 1  \\
	&+ 2N\exp ( { - N{{\sin }^2}\varphi } )\sin \left[ {\frac{N}{2}\sin \left( {2\varphi } \right)} \right]\sin \theta.
	\label{e19} 
	\end{align}
	
	Based on Eqs. (\ref{e16}) and (\ref{e19}), the optimal sensitivity of phase $\varphi$ is given by
	\begin{equation}
	\min \left[ {\frac{{\sqrt {\left\langle {\hat X_B^2} \right\rangle  - {{\left\langle {{{\hat X}_B}} \right\rangle }^2}} }}{{\left| {{{\partial \left\langle {{{\hat X}_B}} \right\rangle } \mathord{\left/
							{\vphantom {{\partial \left\langle {{{\hat X}_B}} \right\rangle } {\partial \varphi }}} \right.
							\kern-\nulldelimiterspace} {\partial \varphi }}} \right|}}} \right] = \frac{1}{{{N^{{3 \mathord{\left/
							{\vphantom {3 2}} \right.
							\kern-\nulldelimiterspace} 2}}}}}.
	\label{e20}
	\end{equation}
	Equation (\ref{e20}) manifests that the sensitivity gets its optimal value $\delta \varphi _{\rm min}={{{N^{{-3 \mathord{\left/{\vphantom {3 2}} \right. \kern-\nulldelimiterspace} 2}}}}}$ when the conditions $\varphi=0$ and $\theta=\pi/2$ are satisfied.
	It should be noted that here $\theta=\pi/2$ is the solution of   equation ${{ {{{\partial  {\theta} } \mathord{\left/
						{\vphantom {{\partial \left\langle {\theta} \right\rangle } {\partial \varphi }}} \right.
						\kern-\nulldelimiterspace} {\partial \varphi }}}}} = 0 $.
	The relationship between $\varphi$ and $\theta$ is given by Eq. (\ref{e16}) \cite{Zhang18}.

\begin{figure}[htbp]
		\centering\includegraphics[width=0.48\textwidth]{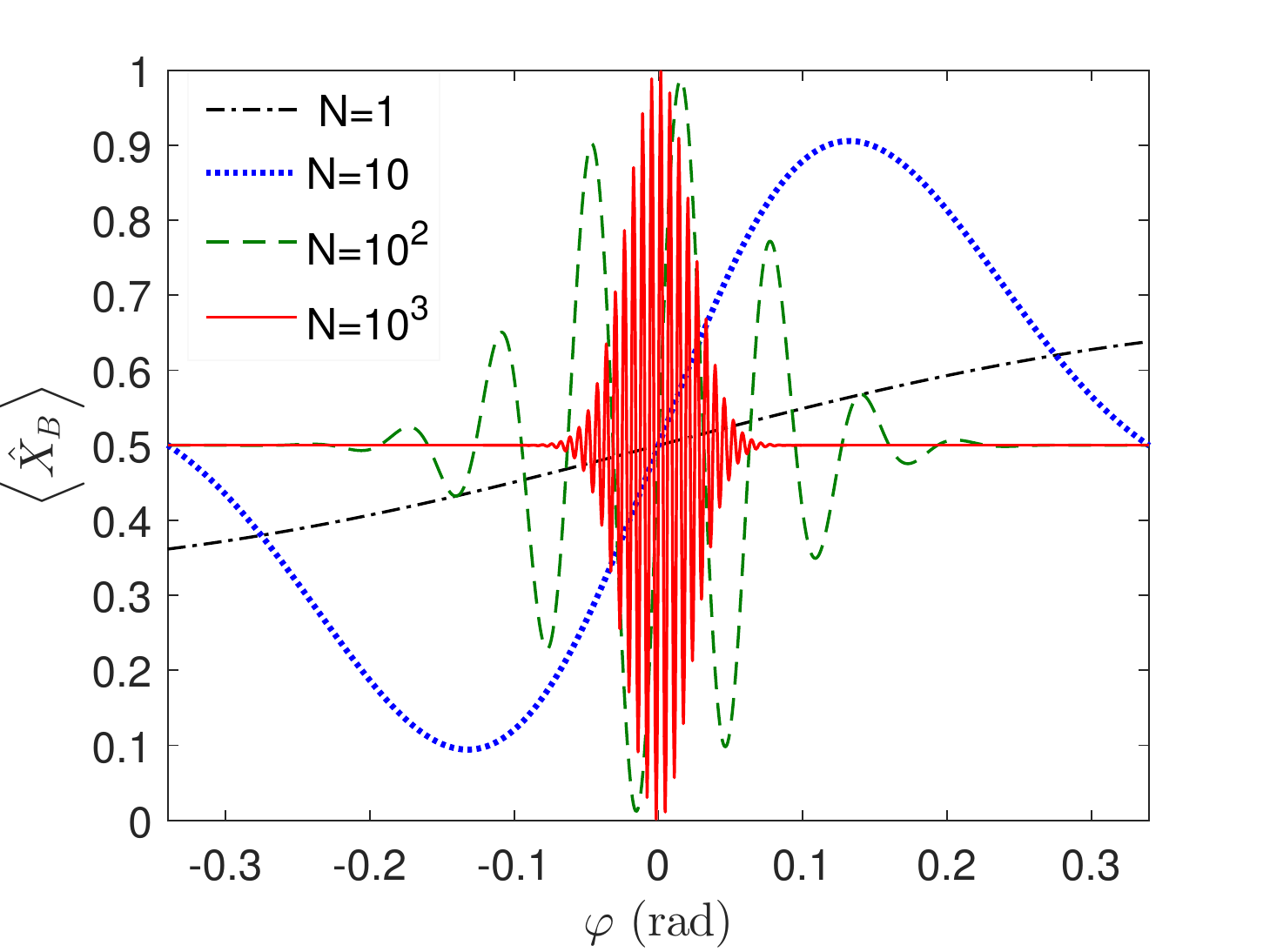}
		\centering\includegraphics[width=0.48\textwidth]{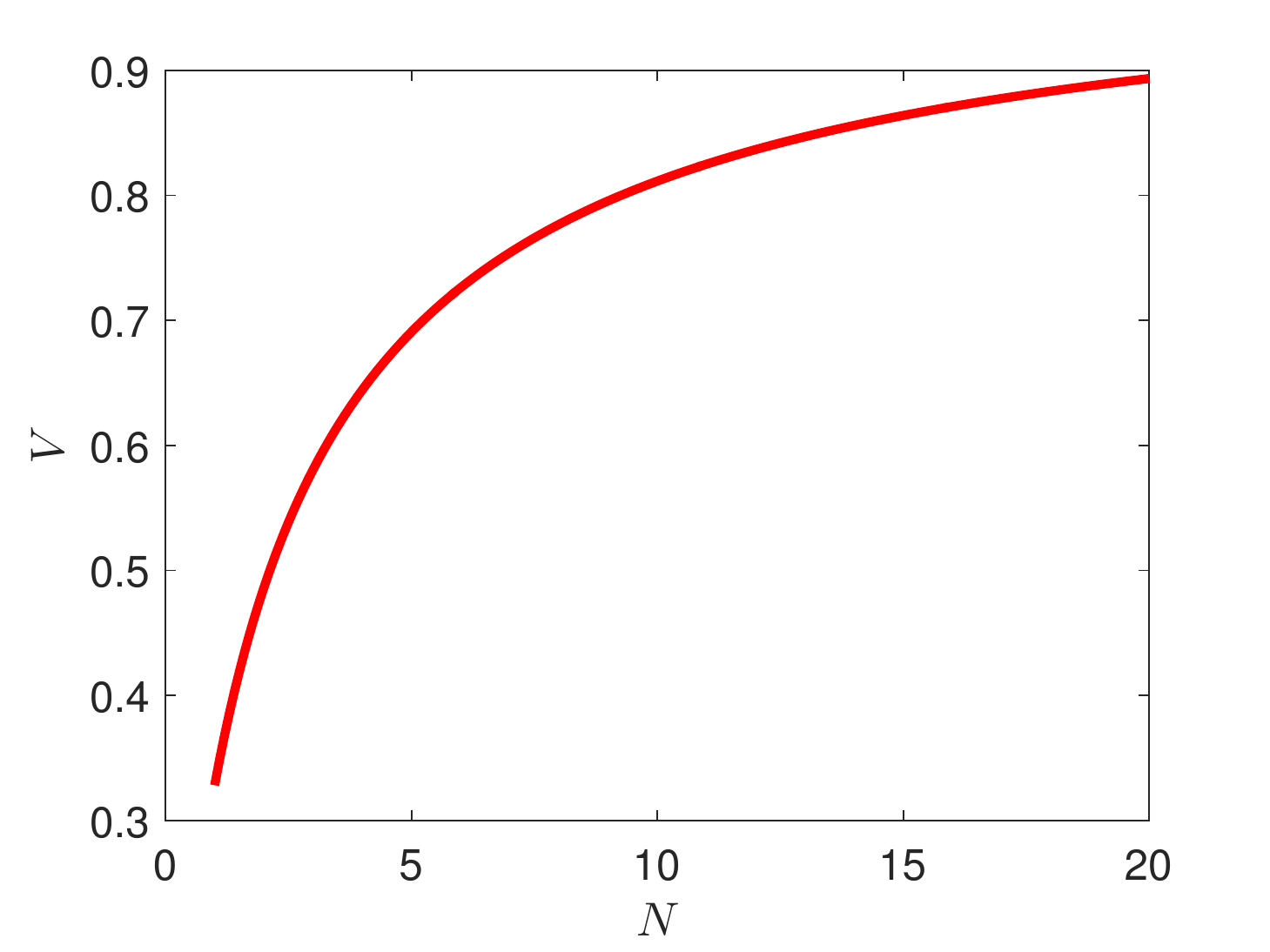}
		\caption{(a) The normalized expectation value against the nonlinear phase shift. (b) The visibility of the expectation value against the mean photon number.}
		\label{f3}
	\end{figure}

	To observe the behavior of the expectation value, in Fig. \ref{f3}(a) we plot the normalized expectation value as a function of the nonlinear phase shift.
	Figure \ref{f3}(a) suggests that the full width at half maximum gets narrow with the increase of the mean photon number; meanwhile, there exist multi-fold oscillating fringes in an envelope.
	These narrow fringes originate mainly from the exponential term $\exp ( { - N{{\sin }^2}\varphi } )$.
	The envelope is modulated by the sine term $\sin [ {{{{N {\sin ( {2\varphi })}}/{2}}}}]$, and the oscillation corresponds to the term $N\sin \left( {2\varphi } \right) $ in the sine function.
	According to the definition of visibility \cite{dowling2008quantum}
	\begin{equation}
	V = \frac{{{{\left\langle {{{\hat X}_B}} \right\rangle }_{\max }} - {{\left\langle {{{\hat X}_B}} \right\rangle }_{\min }}}}{{ \left|{{{\left\langle {{{\hat X}_B}} \right\rangle }_{\max }}}\right| +  \left|{{{\left\langle {{{\hat X}_B}} \right\rangle }_{\min }}} \right| }},
	\label{e21}
	\end{equation}
	in Fig. \ref{f3}(b) we give dependence of the visibility of our protocol on the mean photon number.
	With increasing the photon number, the visibility increases at a quick rate.
	We can get the expectation value of which the visibility is in exceed of 90\% so long as the number of photons is greater than 20.

	\section{Fundamental sensitivity limit}
	\label{s4}
	In the last section, we have calculated the sensitivity of our protocol.
	For evaluating the optimality of measurement strategy, in this section, we give the QFI and compare it with our sensitivity.
	Of the linear phase estimation, the QFI-only calculation may loosen the tightness of sensitivity limit.
	For SU(2) interferometers,  using operators $\exp [ i\varphi ({{\hat a}^\dag }\hat a - {{\hat b}^\dag }\hat b) / 2 ]$ and $\exp ( {i\varphi{{\hat a}^\dag }\hat a } )$ to describe the estimated phase $\varphi$, one may get two different QFI with respect to the same inputs.
	Consider a coherent state $\left| \alpha  \right\rangle $ and a vacuum as inputs, the QFI of the former is $N$ and that of the latter is $2N$ \cite{PhysRevA.85.011801,PhysRevA.96.052118}.		
	Regarding SU(1,1) interferometers, there exist a large number of  similar situations.
	With the same input states, two different QFI may be obtained if one uses operators $\exp ( {i\varphi{{\hat a}^\dag }\hat a } )$ and $\exp ( {i\varphi{{\hat b}^\dag }\hat b } )$ to describe the phase shift \cite{PhysRevA.99.042122}.
	That is, two different sensitivity limits may be obtained from the same physical configurations.
	To circumvent this overestimation, some studies capitalize on the phase-averaging approach to calculate the QFI; accordingly, this problem is partially resolved by this approach in both SU(2) and SU(1,1) interferometers.
	The detailed discussions can be found in Refs. \cite{PhysRevA.85.011801,PhysRevA.96.052118,PhysRevA.99.042122}.
	
	Due to the above reasons, here we deploy the phase-averaging approach to calculate the QFI.
	For our protocol, the density matrix for the input state can be written as
	\begin{equation}
	{\rho _{\rm in}} = \sum\limits_{p,q = 0}^\infty  {{s_{pq}}\left| p \right\rangle \left\langle q \right|}  \otimes \left| 0 \right\rangle \left\langle 0 \right|
	\end{equation}
	in twin Fock basis, where ${s_{pq}} ={{e^{ - N}}{{\left| \alpha  \right|}^{p + q}}} /{\sqrt {p!q!} } $. 
	According to the phase-averaging approach, we need to erase the phase reference information, and then the input turns to a mixed state from a pure state,
	\begin{align}
	\nonumber \bar \rho  &= \int_0^{2\pi } {\frac{{d\phi }}{{2\pi }}} \exp ( { i\phi {{{\hat a}^\dag }\hat a }} ){\rho _{\rm in}}\exp ( { - i\phi {{{\hat a}^\dag }\hat a }})\\
	&= \sum\limits_{p = 0}^\infty  {{{ {{s_{pp}}} }}\left| p \right\rangle \left\langle p \right|}  \otimes \left| 0 \right\rangle \left\langle 0 \right|.
	\end{align}
	After the first beam splitter, this density matrix evolves into
	\begin{equation}
	\rho ' = {U_{\rm BS}}\bar \rho U_{\rm BS}^\dag = \sum\limits_{p = 0}^\infty  {{{{{s_{pp}}} }}\left| {{\psi _p}} \right\rangle \left\langle {{\psi _p}} \right|}
	\end{equation}
	with the state
	\begin{equation}
	\left| {{\psi _p}} \right\rangle  = \sum\limits_{j = 0}^p {\sqrt {\frac{{p!}}{{j!\left( {p - j} \right)!}}} } {\left( {\frac{1}{{\sqrt 2 }}} \right)^p}\left| j \right\rangle  \otimes \left| {p - j} \right\rangle. 
	\label{D6}
	\end{equation}
	The above state is a pure state and obeys the orthogonality 
	$\left\langle {\psi _{p'}}
	{\left | {\vphantom {\psi _{p'} \psi _{p}}}
		\right. \kern-\nulldelimiterspace}{\psi _{p}} \right\rangle  = {\delta _{p'p}}$.
	Related to this, the relationship between total QFI and the QFI of the state in Eq. (\ref{D6}) is given by
	\begin{align}
	{{\cal F}_{\varphi }} &= \sum\limits_{p = 0}^\infty  {{{ {{s_{pp}}} }}{\cal F}_{ \varphi }^p}. 
	\label{D11}
	\end{align}
	
	In order to calculate ${\cal F}_{\varphi }^p$, we give normal order of the estimator ${{\hat O}_\varphi} = {{\hat a}^{\dag 2}}{{\hat a}^2}$ and that of its square $\hat O_\varphi^2 ={{\hat a}^{\dag 4}}{{\hat a}^4}  {\rm + }4{{\hat a}^{\dag 3}}{{\hat a}^3} {\rm + } 2{{\hat a}^{\dag 2}}{{\hat a}^2}$.
	For the state $\left| {{\psi _p}} \right\rangle $, the expectation value of normal order can be calculated by 
	\begin{equation}
	\left\langle {{{\hat a}^{\dag {m}}}{{\hat a}^m}} \right\rangle  =\mathop \prod \limits_m \frac{1}{{{2^m}}}\frac{{p!}}{{\left( {p - m} \right)!}}.
	\label{D13}
	\end{equation}	
	Further, the QFI of the state $\left| {{\psi _p}} \right\rangle $ is equal to
	\begin{align}
	{\cal F}_{\varphi }^p &= 4\left( {\left\langle {\hat O_\varphi^2} \right\rangle  - {{\left\langle {{{\hat O}_\varphi}} \right\rangle }^2}} \right)= \frac{1}{2}p\left( {p - 1} \right)\left( {2p - 1} \right).
	\label{D12}
	\end{align}
	Combining Eqs. (\ref{D11}) and (\ref{D12}), the total QFI can be expressed as
	\begin{align}
	{{\cal F}_{\varphi}} &= {e^{ - N}}\sum\limits_{p = 0}^\infty  {\frac{{{N^p}}}{{2\left( {p - 2} \right)!}}} \left( {2p - 1} \right).
	\end{align}
	The corresponding sensitivity limit is calculated via the equation $\delta \varphi = { {\cal F}_{\varphi }^{-1}}$.
	
\begin{figure}[htbp]
		\centering\includegraphics[width=0.48\textwidth]{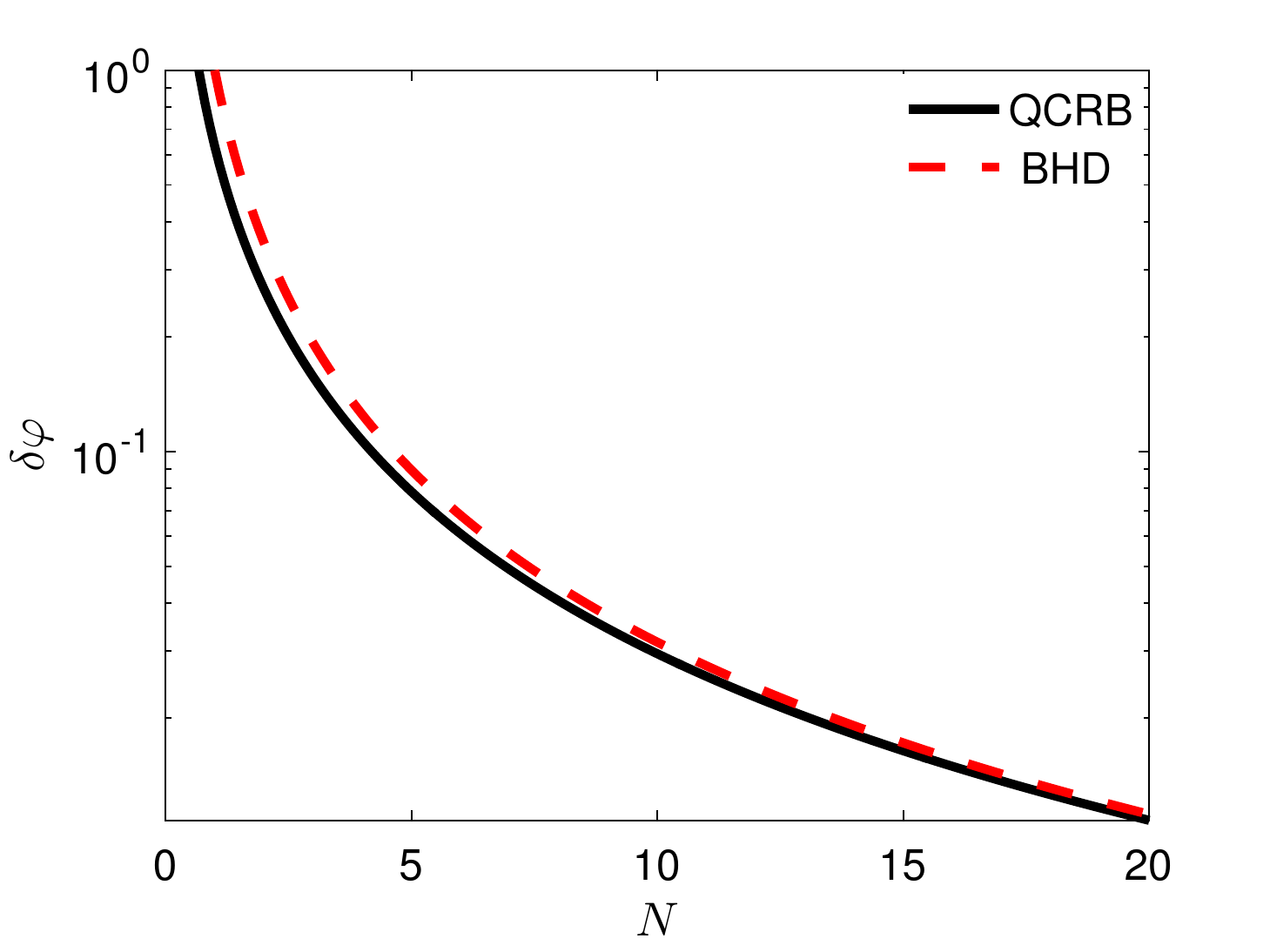}
		\centering\includegraphics[width=0.48\textwidth]{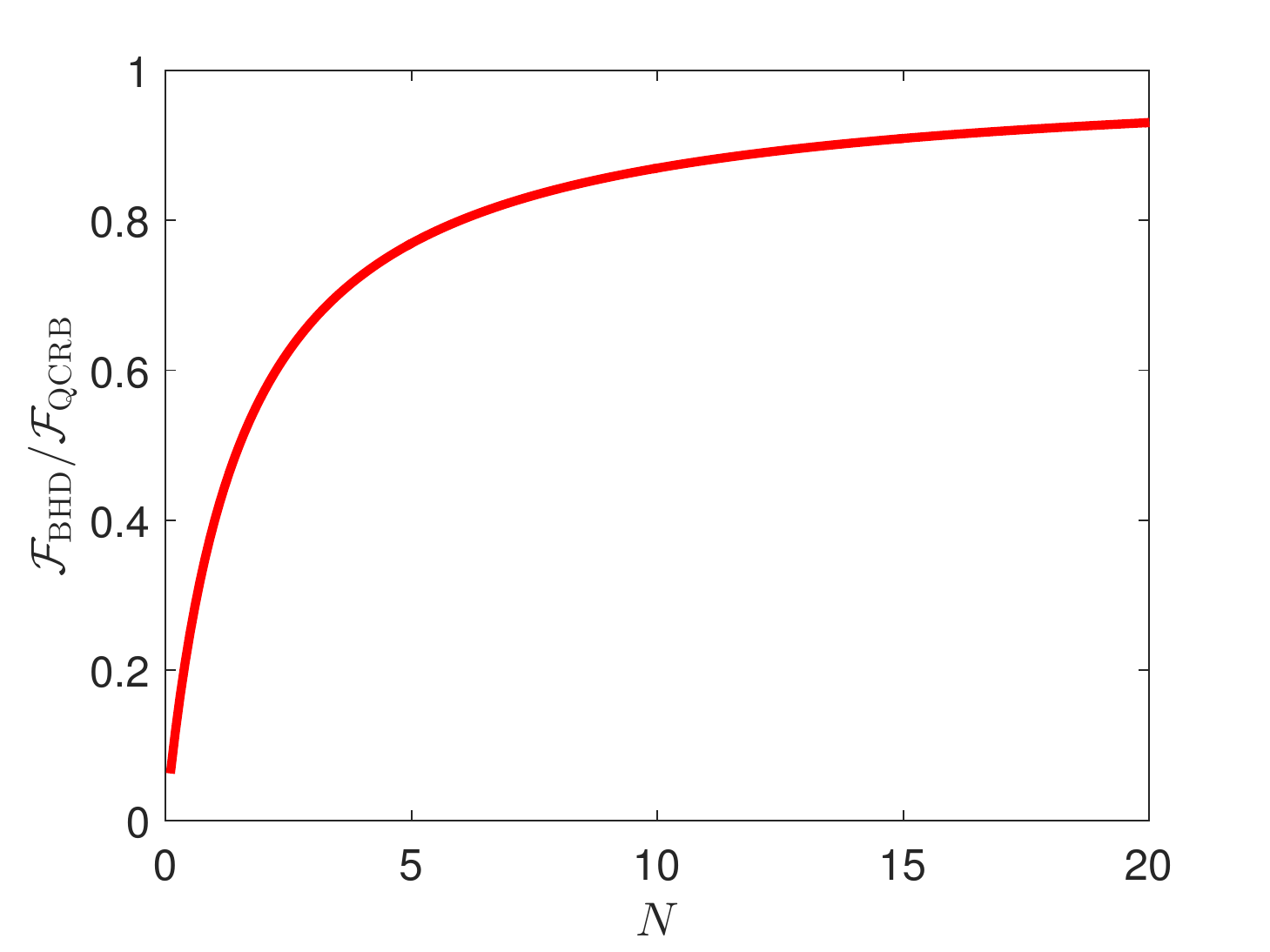}
		\caption{(a) The optimal sensitivity with balanced homodyne detection against the mean photon number. (b) The normalized available Fisher information against the mean photon number.}
		\label{f4}
	\end{figure}

	Figure \ref{f4}(a) demonstrates the quantum Cram\'er-Rao (QCR) bound—inverse square root of the QFI—and the optimal sensitivity obtained by our protocol, respectively.
	For the region of small photon number, the sensitivity is slightly inferior to the bound.
	With further increasing the number of photons, the sensitivity approaches the QCR bound.
	This reveals that balanced homodyne detection is a nearly optimal strategy.
	In Fig. \ref{f4}(b), we present the normalized available Fisher information, ${\cal F}_{\rm BHD}/{\cal F}_{\rm QCRB}$.
	This quotient describes the degree of optimality, only when the optimal strategy is performed does the quotient sit at 1.
	From the figure we can find that balanced homodyne detection continuously approaches the optimal strategy with increasing the mean photon number.
	
	\section{Lossy effect on estimation protocol}
	\label{s5}
	In general, realistic interferometers need to deal with a trade-off among the optimality, robustness, and complexity.
	For our protocol, there is needless to excessively anxious about the complexity, since a conventional interferometer can be competent.
	Regarding the optimality, we have proved that balanced homodyne detection is a nearly optimal measurement strategy, which approaches the QCR bound.
	Thus, in this section we briefly discuss the robustness of our protocol, the effect of photon loss on the optimal sensitivity is considered.

	For simplicity, we merely discuss the same losses in the two paths.
	Photon loss is usually modeled by inserting a fictitious beam splitter with transmissivity $T$ and reflectivity $L$. 
	For a single-mode state, the reflected photons enter the surrounding thermal bath, known as photon loss.
	The first scenario we consider is the photon loss before the nonlinear phase shift, at this point the amplitude $\alpha$ of the coherent state becomes $T\alpha$.
	As a consequence, the expectation value of $X$ quadrature reduces to
	\begin{align}
	{\left\langle {{{\hat X}_B}} \right\rangle _{\rm 1}} =  - \left| {\sqrt T \alpha } \right| \left \{ { \exp ( { - N'{{\sin }^2}\varphi })\sin \left[ {\frac{N'}{2}\sin \left( {2\varphi } \right)} \right]} + \sin \theta \right\}
	\label{e31}
	\end{align}
	and that of its square is expressed as
	\begin{align}
	\nonumber {\left\langle {\hat X_B^2} \right\rangle _{\rm 1}} = &- \frac{N'}{2} \left\{ \exp ( { - 2N'{{\sin }^2}\varphi })\cos \left[ {N'\sin \left( {2\varphi } \right)} \right] + \cos \left( {2\theta } \right) \right\} + N' + 1\\ 
	&  + 2N'\exp ( { - N'{{\sin }^2}\varphi } )\sin \left[ {\frac{N'}{2}\sin \left( {2\varphi } \right)} \right]\sin \theta
	\label{e32}
	\end{align}
	with $N'=TN$.
	Based on above calculation results, we have the optimal sensitivity, $\delta \varphi _{\rm 1}=1/(TN)^{-3/2}$, which equals the sensitivity of a lossless interferometer fed by a coherent state with the mean photon number $TN$.
	
	Similarly, for the second scenario—photon loss after the nonlinear phase shift—we get the expectation values,
	\begin{align}
	{\left\langle {{{\hat X}_B}} \right\rangle _{\rm 2}} &= \sqrt T \left\langle {{{\hat X}_B}} \right\rangle, 
	\label{e33}\\
	{\left\langle {\hat X_B^2} \right\rangle _{\rm 2}} &= T\left\langle {\hat X_B^2} \right\rangle  + 1 - T.
	\label{e34}
	\end{align}
	Using Eqs. (\ref{e33}), (\ref{e34}) and error propagation, the optimal sensitivity $\delta \varphi _{\rm 2}=1/(TN)^{-3/2}$ is obtained. 
	An interesting phenomenon is that the optimal sensitivities of the two scenarios are the same, although the sensitivities are not equal for $\varphi \ne 0$.
	Therefore, with respect to the optimal sensitivity, the places where the photon loss occurs are unimportant, $\delta \varphi _{\rm 1}=\delta \varphi _{\rm 2}$.
	
\begin{figure}[htbp]
		\centering\includegraphics[width=0.48\textwidth]{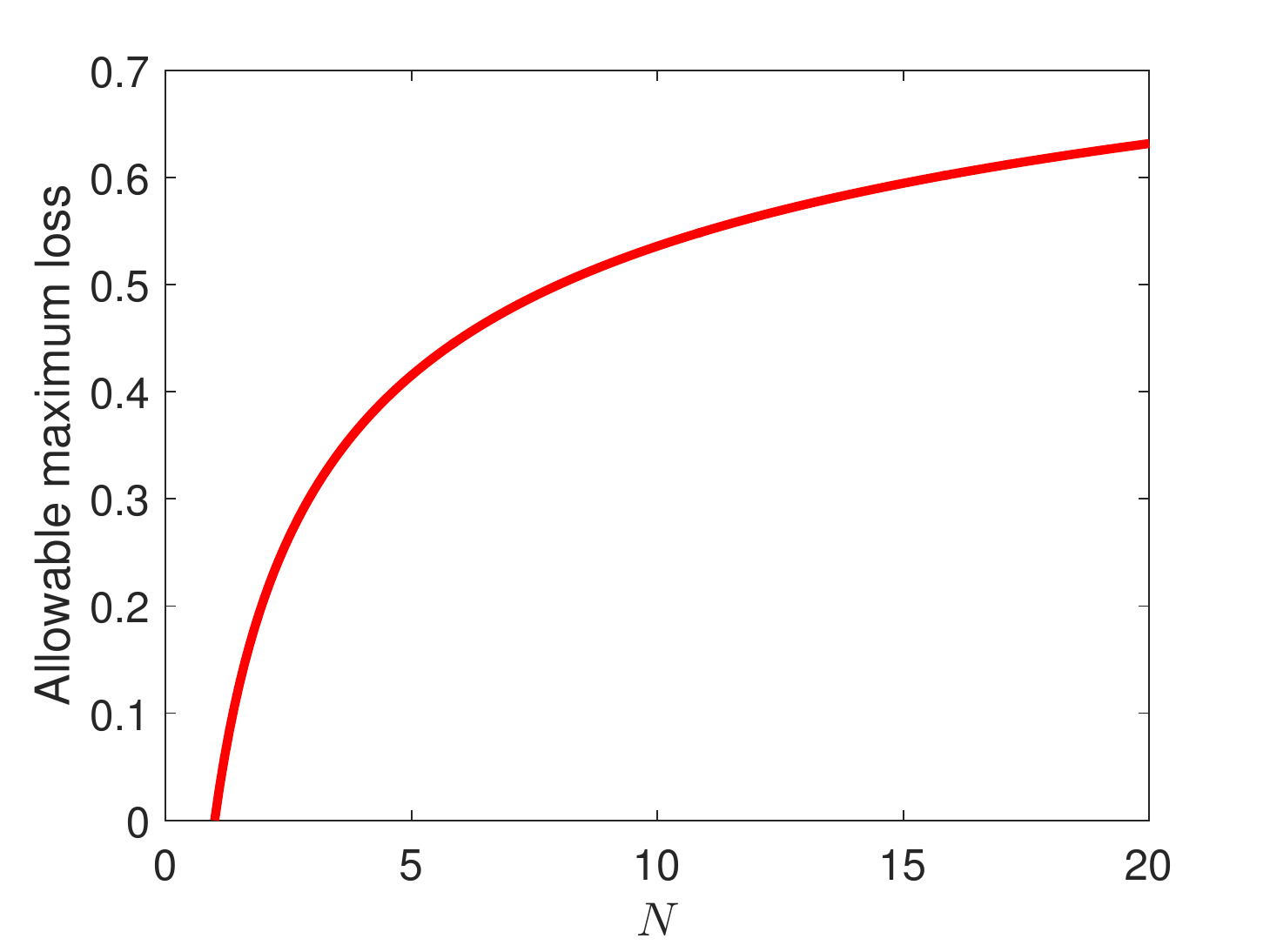}
		\caption{The allowable maximum loss against the mean photon number.}
		\label{f5}
	\end{figure}

	Under the scenario of photon loss, only if the lossy ratio is less than
	$1 {\rm -} {N^{{{ - 1} \mathord{\left/{\vphantom {{ - 1} 3}} \right.\kern-\nulldelimiterspace} 3}}}$ can the sensitivity break the Heisenberg limit, and here $1 {\rm -} {N^{{{ - 1} \mathord{\left/{\vphantom {{ - 1} 3}} \right.\kern-\nulldelimiterspace} 3}}}$ is called as the allowable maximum loss.
	In Fig. \ref{f5}, we describe the relationship between the allowable maximum loss and mean photon number.
	It can be seen that the allowable maximum loss increases rapidly with the increase of mean photon number.
	The protocol can achieve sub-Heisenberg-limited sensitivity and withstand the photon loss in exceed of 60\% for $N=20$.
	This implies that our protocol is robust, and the robustness can be improved by increasing the photon number. 
	
	\section{Conclusion}
	\label{s6}
	This paper focuses on an estimation protocol for the second-order nonlinear phase shifts.
	The input is a coherent state combined with a vacuum, and balanced homodyne detection is performed onto one of the two outputs. 
	In a lossless scenario, we get sub-Heisenberg-limited sensitivity scaling of $N^{-3/2}$, and the output visibility is superior to 90\% with $N \geqslant 20$.
	By taking advantage of the phase-averaging approach, we rule out the virtual component of the QFI brought by hidden resources, and ascertain the fundamental sensitivity limit.
	Compared with this fundamental limit, the sensitivity of our protocol is approximately saturated.
	As a realistic consideration, photon loss is discussed in two scenarios, before and after the nonlinear phase shift.
	The results point out that the optimal sensitivities of these two scenarios are the same; furthermore, the effect of photon loss on the sensitivity is not serious. 
	For the region of $N \geqslant 20$, our protocol stands up to the photon loss in exceed of 60\% and, meanwhile, achieves the Heisenberg limit.
	Overall, our protocol is of approximate optimality and robustness for nonlinear phase estimation.

%

\end{document}